\documentclass[5p,twocolumn]{elsarticle}

\journal{Astroparticle Physics}

\usepackage{graphicx}% Include figure files
\usepackage{epsfig}
\usepackage{amsmath}
\usepackage{dcolumn}% Align table columns on decimal point
\usepackage{bm}% bold math
\usepackage{dcolumn}
\usepackage{color}
\usepackage{verbatim}
\usepackage{paralist}
\usepackage{amssymb}
\usepackage{url}

\usepackage{lineno}
\makeatletter
\def\ps@pprintTitle{%
 \let\@oddhead\@empty
 \let\@evenhead\@empty
 \def\@oddfoot{}%
 \let\@evenfoot\@oddfoot}
\makeatother

\begin{document}

\begin{frontmatter}

\title{ Seasonal Modulation of the $^7$Be Solar Neutrino Rate in Borexino}

\author[GSSI]{M.~Agostini}
\author[Munchen]{K.~Altenm\"{u}ller}
\author[Munchen]{S.~Appel}
\author[Kurchatov]{V.~Atroshchenko}
\author[Milano]{D.~Basilico}
\author[Milano]{G.~Bellini}
\author[PrincetonChemEng]{J.~Benziger}
\author[Hamburg]{D.~Bick}
\author[LNGS]{G.~Bonfini}
\author[Kurchatov]{L.~Borodikhina}
% ...

\cortext[cc]{Corresponding author: spokeperson-borex@lngs.infn.it}
\fntext[Canfranc]{Also at: Laboratorio Subterr\'aneo de Canfranc, Paseo de los Ayerbe S/N, 22880 Canfranc Estacion Huesca, Spain}
\fntext[DIEGO]{Present address: Physics Department, University of California, San Diego, CA 92093, USA}
\fntext[Queens]{Present address: Physics Department, Queen's University, Kingston ON K7L 3N6, Canada}

\address{\bf{The Borexino Collaboration}}

% ...
\address[Hamburg]{Institut f\"ur Experimentalphysik, Universit\"at Hamburg, 22761 Hamburg, Germany}
% ...
\address[Kurchatov]{National Research Centre Kurchatov Institute, 123182 Moscow, Russia}
% ...
\address[LNGS]{INFN Laboratori Nazionali del Gran Sasso, 67010 Assergi (AQ), Italy}
\address[Milano]{Dipartimento di Fisica, Universit\`a degli Studi e INFN, 20133 Milano, Italy}
% ...
\address[PrincetonChemEng]{Chemical Engineering Department, Princeton University, Princeton, NJ 08544, USA}
% ...
\address[Munchen]{Physik-Department and Excellence Cluster Universe, Technische Universit\"at  M\"unchen, 85748 Garching, Germany}
% ...
\address[GSSI]{ Gran Sasso Science Institute (INFN), 67100 L'Aquila, Italy}
% ...

\begin{abstract}

We present the evidence for the seasonal modulation of the $^7$Be neutrino interaction rate with the Borexino detector at the Laboratori Nazionali del Gran Sasso in Italy. 
The period, amplitude, and phase of the observed time evolution of the signal are 
consistent with its solar origin, and the absence of 
an annual modulation is rejected at 99.99\% C.L. The data are analyzed using three methods: the analytical fit to event rate, the Lomb-Scargle and the Empirical Mode Decomposition techniques, which all yield results in excellent agreement.

\end{abstract}


\end{frontmatter}

\section{Introduction}
\label{sec:intro}

Since 2007 Borexino \cite{bib:detector} has measured the fluxes of low-energy neutrinos, most notably those emitted in nuclear fusion reactions and $\beta$ decays along the {\it pp}-chain in the Sun. Borexino was the first experiment to make spectroscopic and real-time measurements of solar neutrinos with energy $<$3 MeV, i.e. below the endpoint energy of long-lived, natural $\beta$ radioactivity: $^{40}$K and the $^{232}$Th and $^{238}$U decay chains.
The detector has made first direct observations of $^7$Be \cite{bib:be7}, {\it pep} \cite{bib:pep}, and {\it pp} \cite{bib:bxpp} solar neutrinos, lowered the detection threshold for $^8$B solar neutrinos \cite{bib:b8}. These measurements deepen our understanding of Solar Standard Model \cite{bib:carlos} and support the MSW-LMA mechanism of neutrino oscillations.  In addition Borexino has detected anti-neutrinos from the Earth and distant nuclear reactors \cite{bib:geo} and has set a new upper limit for a hypothetical solar anti-neutrinos flux \cite{bib:anti}.

Borexino, located deep underground (3,800 m water equivalent) in Hall C of the Gran Sasso Laboratory (Italy), measures solar neutrinos via their interactions with a target of 278 ton organic liquid scintillator.
The ultrapure liquid scintillator (pseudocumene (1,2,4-trimethylbenzene (PC)) solvent with 1.5 g/l 2,5-diphenyloxazole (PPO) scintillating solute) is contained inside a thin transparent spherical nylon vessel of 8.5 m diameter. Solar neutrinos are detected by measuring the energy and position of electrons scattered by neutrino-electron elastic interactions. The scintillator promptly converts the kinetic energy of electrons by emitting photons, which are detected and converted into electronic signals (photoelectrons (p.e.)) by 2,212 photomultipliers (PMT) mounted on a concentric 13.7 m-diameter stainless steel sphere (SSS).

The volume between the nylon vessel and the SSS is filled with 889 ton of ultra pure, non scintillating fluid and acts as a radiation shield for external gamma rays and neutrons.
A second, larger nylon sphere (11.5~m diameter) prevents radon and other radioactive contaminants from the PMTs and SSS from diffusing into the central sensitive volume of the detector. The SSS is immersed in a 2,100 ton water \v{C}erenkov detector meant to detect residual cosmic muons \cite{bib:muon}.

Radioactive decays within the scintillator form a background that can mimic neutrino signals. During detector design and construction, a significant effort was made to minimize the radioactive contamination of the scintillator and of all detector components in contact with it. A record low scintillator contamination of $<10^{-18}$ g/g was achieved for $^{238}$U and $^{232}$Th.

The identification of different components of the solar neutrino flux relies on fitting the recorded energy spectrum with a combination of identified radioactive background components and of solar neutrino-induced electron recoil spectra. The neutrino-induced spectra are derived from Standard Solar Model neutrino energy distributions (SSM \cite{bib:serenelli16}) and include the effect of neutrino oscillation.
The solar origin of the detected neutrinos is determined by the identification of crisp spectral signatures as predicted by the SSM. Exemplary is the Compton-like energy spectrum of electrons scattered by the mono-energetic $^7$Be solar neutrinos. Remarkably, the $^7$Be-induced Compton 'shoulder' was clearly identified with just one month of data \cite{bib:be2007}, thanks to the extremely low radioactive background in the scintillator.

In contrast with water \v{C}erenkov detectors, Borexino cannot retain directional information of individual events due to the isotropic emission of scintillation light; direct solar imaging with neutrinos is thus not possible. The eccentricity of the Earth's orbit, however, induces a modulation of the detected solar neutrino interaction rate proportional in amplitude to the solid angle subtended by the Earth with respect to the Sun (neglecting neutrino oscillation effects). The effect appears as a 6.7\% peak-to-peak seasonal amplitude modulation, with a maximum at the perihelion. Evidence for such a yearly modulation of the $^7$Be signal was already observed with Borexino Phase-I data (collected from May 2007 to May 2010) \cite{bib:long}.
The period and phase were found to be consistent with a solar origin of the signal. 

Yearly modulation searches have also been carried out by other solar neutrino experiments: in particular SNO \cite{bib:SNO} and Super-Kamiokande \cite{bib:SK} found evidence for an annual flux modulation in their time series datasets.
Similar analyses were also performed aiming to search for time-dependencies of solar neutrino rates with periods other than one year. An apparent anti-correlation with solar cycles was suggested by data from the Homestake chlorine experiment \cite{bib:homestake}, and claims of such a periodicity were reported for Super-Kamiokande-I  \cite{bib:mils,bib:stu,bib:stu2}.
The SNO \cite{bib:SNO}, Super-Kamiokande \cite{bib:SK1}, and Gallex/GNO \cite{bib:GNO} collaborations looked for these time variations, but found none in their data.

Here we report an improved measurement of time periodicities of the $^7$Be solar neutrino rate based on 4 years of Borexino Phase-II data, acquired between December 2011 and December 2015.
Borexino Phase-II began immediately after an extensive period of scintillator purification. 
Borexino Phase-II,  in addition to higher statistics, lower background levels and an improved rejection of alpha-decay background, is characterized by the absence of major scintillator handling and thus displays a high degree of stability of the detector, crucially important for identifying time dependent signals. 
In the Borexino Phase-I analysis we based our annual modulation search on the well-established Lomb-Scargle approach as well as on the more recent Empirical Mode Decomposition (EMD) technique. The virtue of the latter technique is its sensitivity to transient modulations embedded in time series, emerging from analyzing data features with more than just standard reference sinusoidal functions.

The analysis reported here analyzes the Borexino Phase-II dataset, described in Sec.~\ref{sec:DataSet}, by employing both the Lomb-Scargle and an updated version of the EMD techniques. Two independent sections of this paper describe the methods of each approach and their respective results (Sec.~\ref{sec:lomb} and Sec.~\ref{sec:EMD}).
For completeness, we have also carried out a search of the annual modulation directly in the time domain, using a straightforward analytical fit (Sec.~\ref{sec:fit}).
All analysis methods clearly confirm the presence of an annual modulation of the $^7$Be solar neutrino interaction rate in Borexino and show no signs of other periodic time variations. 

\section{The data set}

\label{sec:DataSet}

The data of Borexino Phase-II are used for this analysis (1456 astronomical days of data). 
Compared to Borexino Phase-I, background levels have been substantially reduced by an extensive purification campaign that took place during 2010 and 2011. 
Of particular importance for this study is the reduction of the $\hbox{}^{85}$Kr and $\hbox{}^{210}$Bi concentrations, both backgrounds in $^7$Be region.
Data taking has seen only occasional, minor interruptions due to detector maintenance.

\subsection{Event selection}
\label{sec:datasel}

\begin{figure}[!]
\centering{\includegraphics[width=0.95\linewidth, angle=0]{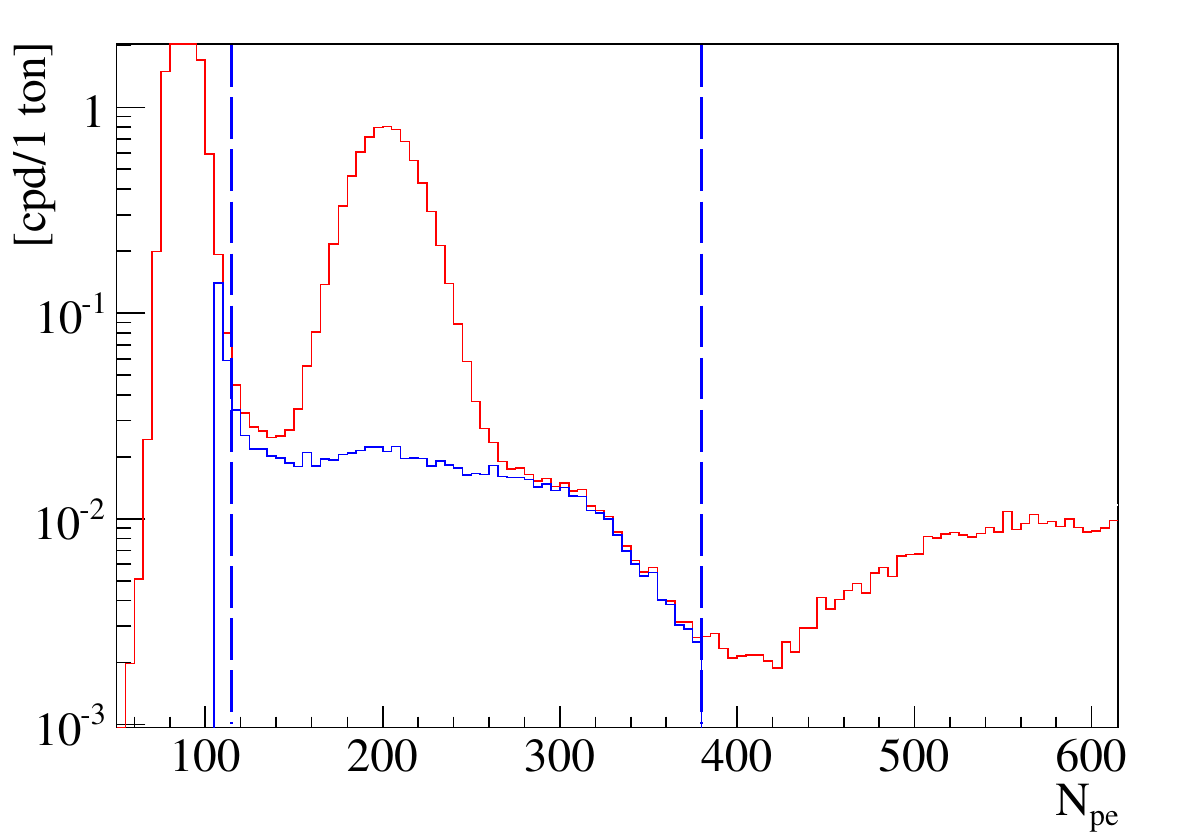}}
\caption{Energy spectrum in number of photoelectrons for events selected in the FV (red line). The blue spectrum are the events with $mlp$ parameter $>0.98$.
A small residue of $^{210}$Po events is still present. The vertical lines define the analysis energy window.}
\label{fig:spectrum}
\end{figure}

A set of cuts described in \cite{bib:long} has been applied on an event-by-event basis to
remove backgrounds and non physical events. In particular, muons and spallation events within 300 ms of parent muons, time-correlated events ($^{214}$Bi-$^{214}$Po), and noise events are identified and removed. In addition, events featuring vertices reconstructed outside a Fiducial Volume (FV) are rejected.
Recoil electrons from the elastic scattering of $^7$Be-$\nu$'s are selected by restricting the analysis to the energy region 
$\sim$215-715 keV ($115-380$ N$_{pe}$). In this range, the major backgrounds are the $\alpha$ decays of $^{210}$Po and the $\beta$ decays of $^{210}$Bi and $^{85}$Kr.
The 5.3 MeV $\alpha$'s appear as a peak at $\sim$450 keV (after quenching) in the energy spectrum (red line in Fig.~\ref{fig:spectrum}). The $\beta$'s
define a continuous spectrum beneath the $^7$Be recoil spectrum (blue line in Fig.~\ref{fig:spectrum}).
The time stability of the background was studied to factor out any influence on the annual modulation search.
Two major changes were implemented for this search from that with Borexino Phase-I data and described below: the FV (Sec.~\ref{sec:vol}) was redefined and an enhanced method for the rejection of $^{210}$Po $\alpha$ background was developed (Sec.~\ref{sec:mlp-psd}).

\subsubsection{Fiducial Volume Selection}
\label{sec:vol}

We define a FV of 98.6~ton by combining a spherical cut of $R=3$~m radius at the center of the detector with two paraboloidal cuts at the nylon vessel poles to reject $\gamma$-rays from the Inner Vessel {\it end-cap} support hardware and plumbing. 

The excluded paraboloids have different dimensions to remove the local background. The paraboloids are defined as $R(\theta)=d/cos^n \theta$, where $\theta$ is the angle with z-axis and $d$ is the distance from the detector center to the paraboloid vertex. The top paraboloid is defined by $d$=250~cm and $n$=12 whitch corresponds to an aperture of $54$~cm of radius;  the bottom one by $d$=-240~cm and $n$=4 which corresponds to a larger aperture of $91$~cm of radius.

\subsubsection{$^{210}$Po Rejection}
\label{sec:mlp-psd}

\begin{figure}[h!]
\begin{center}
\centering{\includegraphics[width = 0.50 \textwidth]{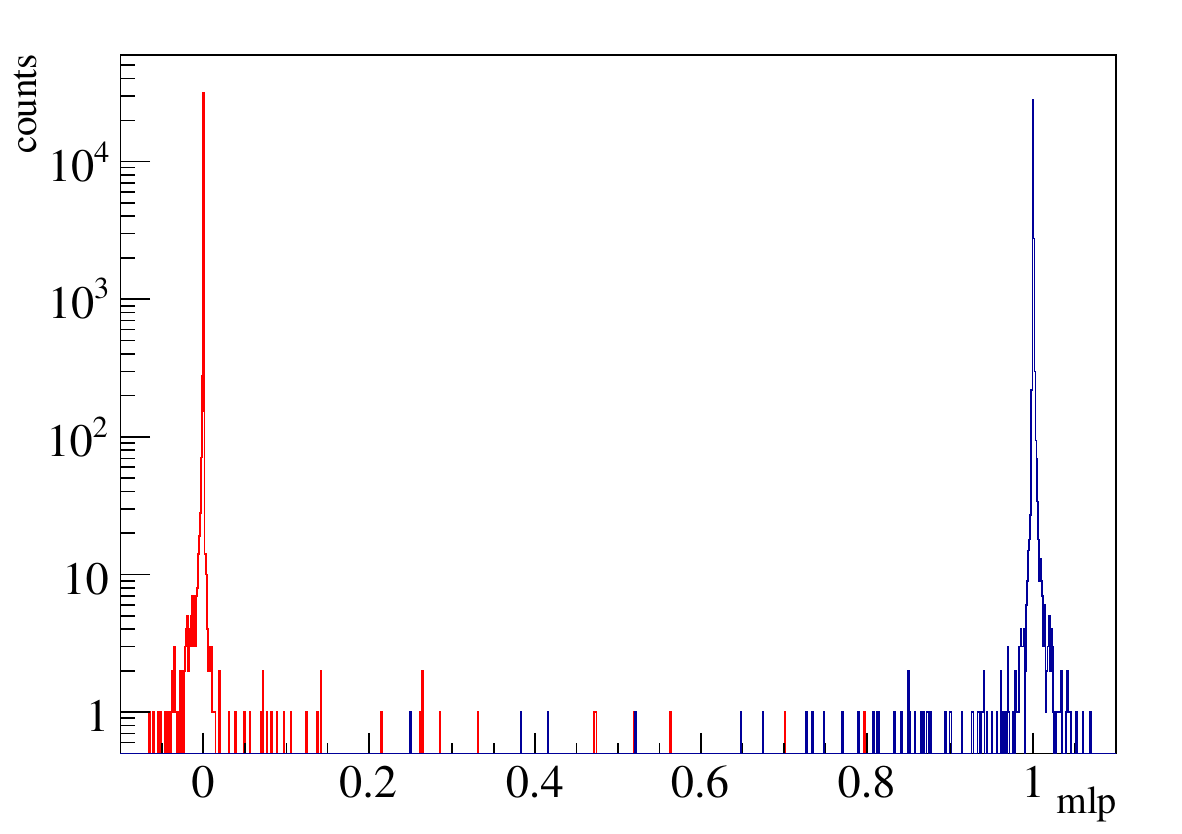}}
\caption{Distribution of $mlp$ variable for $\alpha$ (red) and $\beta$ (blue) events obtained by tagging the $^{214}$Bi-$^{214}$Po time coincidences.}
\label{fig:MLPvar}
\end{center}
\end{figure}

$^{210}$Po in the scintillator constitutes a background for the search of time-varying signals because of its decay half-life of 138 days.
In general $\alpha$-backgrounds and $\beta$-events in a liquid scintillator can be efficiently separated exploiting the largely different shapes of the scintillation pulses~\cite{bib:detector}.
A novel pulse-shape method based on MultiLayer Perceptron (MLP) machine learning algorithm was applied to distinguish between the scintillation pulses of $\alpha$ and $\beta$ particles with high efficiency.
This multivariate method uses a neural network based on 13 $\alpha/\beta$ 
discriminating input variables, that are computed for each event from the time distribution of reconstructed PMT hits. Clean samples of $\alpha$ and $\beta$ events were obtained from the radon daughters $^{214}$Po and $^{214}$Bi to train the neural network. 
The resulting $mlp$ parameter assumes values mostly between 0 ($\alpha$) and 1 ($\beta$). 
Figure \ref{fig:MLPvar} shows the distributions of the $mlp$ parameters for the 
$^{214}$Po and $^{214}$Bi event samples.

The MLP provides excellent $\alpha$-$\beta$ discrimination: with the $mlp$ parameter threshold set at 0.9 to retain $\beta$'s, the $\alpha$ rejection efficiency is $>99.98\%$ for $^{214}$Po candidate events (7.7 MeV). 
The discrimination technique is based upon scintillation pulse shape, therefore we expect a reduced performance for the lower energy $^{210}$Po $\alpha$'s (5.3 MeV) due to lower photoelectron statistics.  
In this case, for a clean $\beta$-like electron-recoil sample, we select events with $mlp >0.98$. 
Fig.~\ref{fig:spectrum} shows the energy spectrum with and without $\alpha$ subtraction (blue and red lines). The small residual $^{210}$Po events and the unaffected $\beta$ spectrum illustrate the efficacy of the discrimination.

\subsection{Residual Background}
\label{sec:res_bkg}

There are two main sources of background for this analysis: the residual $^{210}$Po activity, and the stability of $^{210}$Bi and $^{85}$Kr~$\beta$-decays in the FV. 

\subsubsection{Residual $^{210}$Po}
\label{sec:mlpeff}

At the beginning of Borexino Phase-II (Dec. 2011), the count rate of $~^{210}$Po was  
$\sim 1400\;$ cpd/100~ton. Estimating an $mlp$ $\alpha$-$\beta$ efficiency of $\simeq 99\%$, the residual $\alpha$ contamination of the $\beta$ spectrum is $R_{\alpha} \sim 14$ cpd/100~ton, 
comparable to an average $\beta$ count rate ($\nu$-signal and background) $ \overline{R}_{\beta}\sim 40$ cpd/100~ton distributed over the entire analysis energy region.
We estimated the efficiency of the MLP cut by looking for any exponentially
decaying $^{210}$Po residual still present in the dataset. The residual amount of $R_{\alpha}$ has been subtracted for a given $mlp$ cut in each time bin $R(t)$:

\begin{equation}
\label{eq:mlp_corr}
{R}_{\beta}(t) =R(t)-\xi_{mlp}\cdot R_{\alpha}(t),
\end{equation}
where $\xi_{mlp}$ is the `inefficiency' parameter.

For $\xi_{mlp} = 1\%$  the exponential component due to the residual alphas become negligible in the overall time series of the dataset, leaving the remaining $\beta$'s rates with a constant average value in time.

\subsubsection{Background stability}

The $\beta$-decays of $^{210}$Bi and $^{85}$Kr cannot be distinguished from recoil electrons of the same energies induced by neutrinos.  To study the stability of the background rate over time, we compared the spectral fits to the data divided in short periods.
The fit procedure
is the same as in the $^7$Be analysis~\cite{bib:long}.
No appreciable variation of the background rate is observed within uncertainties.

\subsection{Detector Stability}
The stability of the detector response also needs to be characterized, in particular of energy and position reconstruction and fiducial mass.

\subsubsection{Energy and Position Reconstruction}
The stability of the energy scale over time was checked by comparing the number of events in the selected energy window and in the FV with those expected by Monte Carlo. A detailed simulation that includes the run per run detector performance is used. The stability of the energy scale over the period of interest was proven to be better than 1$\%$, adequate for our purposes. 

\subsubsection{Fiducial Mass}

The liquid scintillator density varies with temperature as:  
$\rho_{PC}=((0.89179\pm 0.00003)-(8.015 \pm 0.009)10^{-4}\times T)$ g/cm$^3$,
where T is the temperature in degrees Celsius \cite {bib:long}.
The temperature is monitored at various positions inside the detector. The volume closest to the IV where temperature is recorded is the concentric Outer Buffer, where the thermal stability is measured to be better than 1$^\circ$~C. 
In the FV, the maximum scintillator mass excursion corresponding to temperature variations is 0.1 ton, $\sim0.1\%$ of the FV mass. 
A Lomb-Scargle analysis (Sec.~\ref{sec:lomb}) on the temperature data was performed. The largest amplitude corresponded to a frequency of $\sim$0.6 year$^{-1}$, reflecting a significant real trend which anyhow cannot mimic  the annual modulation.

\section{Modulation analysis}
\label{sec:anal}

%As reported in \cite{bib:long}, 
We have implemented three alternative analysis approaches to identify the seasonal modulation. The first is a simple fit to the data in the time domain (Sec.~\ref{sec:fit}). The second is the Lomb-Scargle method (Sec.~\ref{sec:lomb})~\cite{bib:Lomb,bib:Scargle}, an extension of the Fourier Transform approach. The third method is the Empirical Mode decomposition (EMD)~(Sec.~\ref{sec:EMD})~\cite{bib:Huang98}.

For each approach we define a set of time bins of equal length $t_k$ and their
corresponding event rate $R(t_k)$, obtained as the ratio of the number of selected events and the corrected life time (subtracted of the muon veto dead time and any down-time between consecutive runs).

The time bins are too short to allow extracting a value of the $^7$Be neutrino interaction rate via a spectral fit. %analysis, due to insufficient statistics for a reliable fit.
We use the raw $\beta$-event rate instead, which include background contributions.

\subsection{Fit to the Event Rate}
\label{sec:fit}

Due to Earth's orbital eccentricity $(\epsilon = 0.0167)$, the total count rate is expected to vary as

\begin{equation}
R(t) = R_0 + \overline{R}\left[1 + \epsilon\cos\frac{2 \pi}{T}\left(t- \phi\right)\right]^2 
\label{eq:mod} 
\end{equation}
where T is the period (one year), $\phi$ is the phase relative to the perihelion, $\overline R$ is the average neutrino interaction rate and $R_0$ is the time independent background rate.  This formalism is consistent with the MSW solution in which are no additional time modulations, at the $^7$Be energies \cite{bib:MSW}.

\begin{figure}[t]
\begin{center}
\centering{\includegraphics[width = 0.50 \textwidth,angle=0]{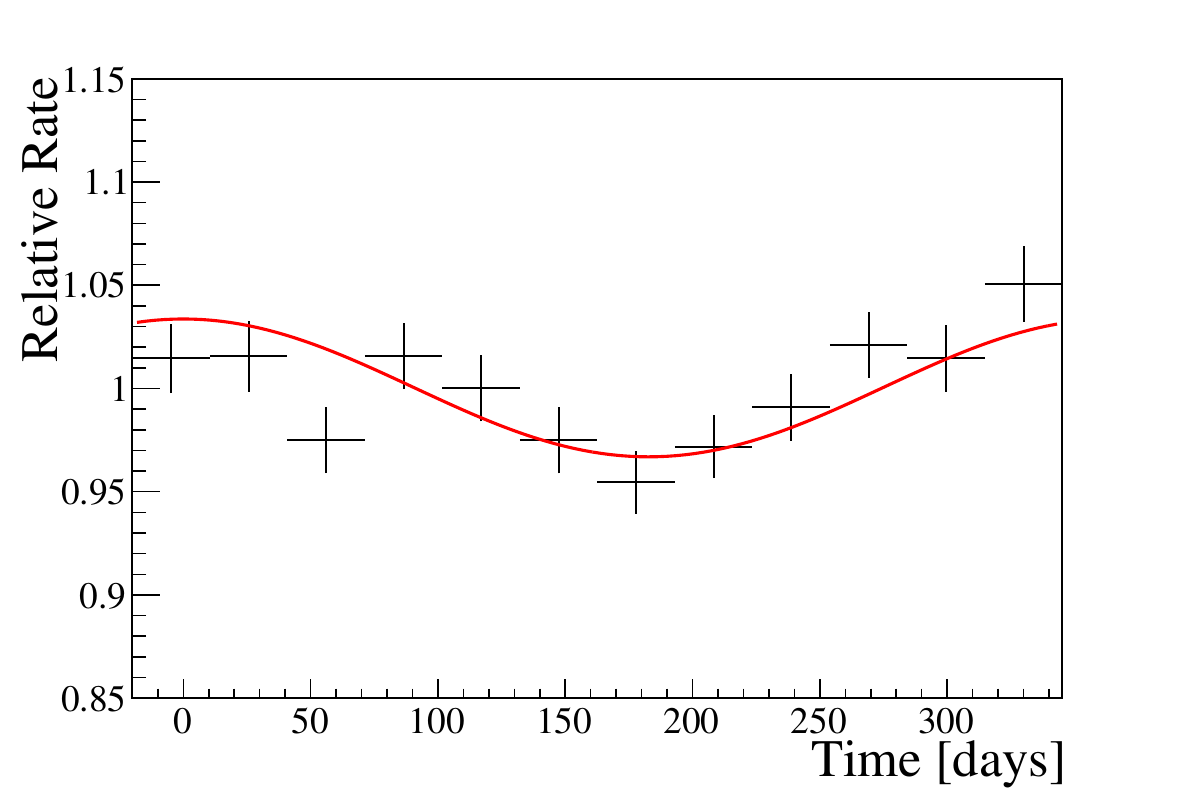}}
\caption{Measured monthly event rate [cpd/100 ton] relative to the average rate of $\beta$-like events passing selection cuts. Data from different years are cumulated. The line is the expected variation according to eq.~\ref{eq:mod}, parametrizing the effect of the Earth's orbit around the Sun. Time bins are 30.43 days long.}
\label{fig:fit_year}
\end{center}
\end{figure}

In this approach, the event rate as a function of the time is fit with the function defined in equation~\ref{eq:mod}.
Figure~\ref{fig:fit_year} shows the folded, monthly event rate relative to the average rate measured in Borexino, with $t=0,\, 365$ representing perihelia. Data from the same months in successive years are added into the same bin.
Having normalized to 1 the overall mean value, the data are compared with Eq.~\ref{eq:mod} and show good agreement with a yearly modulation with the expected amplitude and phase.
The no modulation hypothesis is excluded at  3.91~$\sigma$ (99.99\% C.L.) by comparing the $ \chi^2$ obtained with and without an annual periodicity.  

\begin{figure}[h!]
\begin{center}
\centering{\includegraphics[width = 0.50 \textwidth,angle=0]{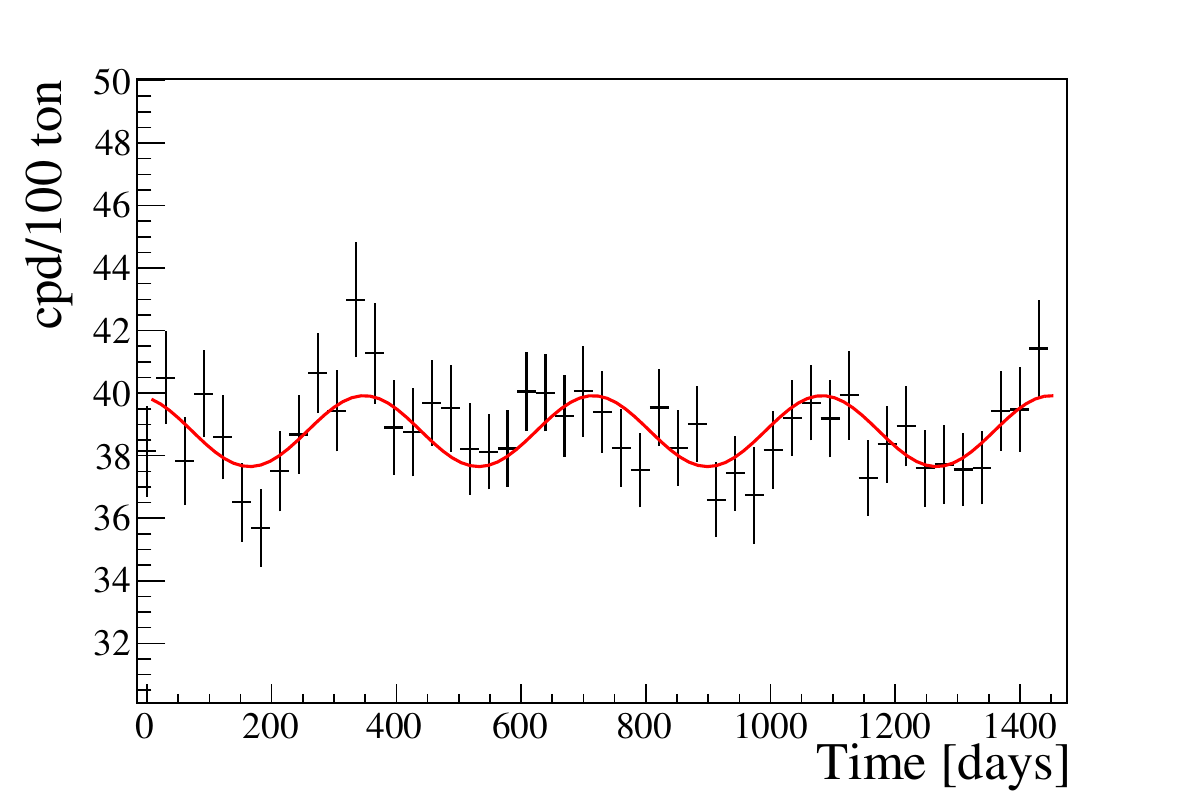}}
\caption{Measured rate of $\beta$-like events passing selection cuts in 30.43-days long bins starting from Dec 11, 2011. The red line is resulting function from the fit with the Eq.~\ref{eq:mod}.}
\label{fig:fit_sin}
\end{center}
\end{figure}

To extract the modulation parameters, we perform a $\chi^2$ fit of the data with 30.43-day bins, without folding multiple years on top of each other.
Figure~\ref{fig:fit_sin} shows the event rate (in cpd/100 ton) along with the best fit.  
From~\cite{bib:be7}, the expected neutrino average rate in this energy range is $\sim$32 \ cpd/100~ton. 
The fit returns an average neutrino rate of $\overline{R}=33 \pm 3$~(cpd/100~ton), within 1$\sigma$ of the expected one ($\chi^2/ndof = 0.68$, $ndof=42$).
The best-fit eccentricity is $\epsilon = 0.0174\pm 0.0045$, which corresponds to an amplitude of the modulation
of $(7.1 \pm 1.9)\%$, and the best-fit period is $T=367\pm 10$ days. Both values are in agreement with the expected values of $6.7\%$ and of $T=365.25$ days. 
The fit returns a phase of $\phi=-18 \pm24$ days.  The robustness of the fit has been studied by varying the
bin size between 7 and 30 days, by shifting the energy range for selected events, and with and without $\alpha-\beta$ $mlp$ inefficiency. 
Fit results are found not to vary greatly and are all in agreement with the expected modulation due to the Earth's orbit 
eccentricity. The resulting systematic uncertainty on the eccentricity is $10\%$.

\subsection{The Lomb-Scargle method}
\label{sec:lomb}

The second approach uses the Lomb-Scargle method. This extension of the Fourier Transform is well suited for our conditions since it can treat data sets that are not evenly distributed in time. 
In the Lomb-Scargle formalism, the Normalized Spectral Power Density, $P(f)$, also known as the Lomb-Scargle periodogram and derived for $N$ data points ($R_1 \dots R_j \dots R_N$) at specific times $t_j$, is 
evaluated and plotted for each frequency $f$ as:

\begin{equation*}
P(f)=\cfrac{1}{2\sigma^2} \left\{ \cfrac{ \left[ \Sigma_j (R_j - \overline{R}) \cos \omega
(t_j-\tau) \right] ^2 }{ \Sigma_j \cos^2\omega (t_j-\tau)} \right.
\end{equation*}
\begin{equation}
\left.+\cfrac{\left[\Sigma_j (R_j -\overline{R})\sin \omega (t_j-\tau)\right]^2}{\Sigma_j \sin^2\omega (t_j-\tau)} \right\}
\label{equ:lomb}
 \end{equation}

% Eq 
\begin{eqnarray*}
\overline{R} = \cfrac{R_1 + R_2 + R_3 + ...+R_N}{N} = \frac{1}{N} \sum_{j=1}^N R_j \\
\sigma^2 = \frac{1}{N-1} \sum_{j=1}^N \left( R_j - \overline{R} \right)^2 \\
\tan 2 \omega \tau = \cfrac{\sum_j \sin 2 \omega t_j}{\sum_j \cos 2 \omega t_j} 
\end{eqnarray*}
where $\omega=2 \pi f$.
After finding the frequency $f_0$ corresponding to the maximum of the Lomb-Scargle Power distribution~\cite{bib:Scargle,bib:gio}, the sine wave that best describes the time-series, in the case of 
a pure signal, is:

\begin{equation}
\label{equ:sinwave}
R(t)=A~cos~ \omega_0 t + B~sin~\omega_0 t
\end{equation}
where, for $\omega_0=2 \pi f_0$ and
\begin{equation*}
A=\cfrac{1}{2\sigma^2} \cfrac{\left[\Sigma_j R_j\cos \omega_0 (t_j-\tau)\right]^2}{\Sigma_j \cos^2\omega_0 (t_j-\tau)} 
\end{equation*}
\begin{equation*}
\quad B=\cfrac{1}{2\sigma^2} \cfrac{\left[\Sigma_j R_j\sin \omega_0 (t_j-\tau)\right]^2}{\Sigma_j \sin^2\omega_0 (t_j-\tau)}
\end{equation*}

The modulation amplitude is the peak-to-peak variation of the curve resulting from Eq.~(\ref{equ:sinwave}).

\begin{figure}[t]
\begin{center}
\centering{\includegraphics[width = 0.50\textwidth, angle=0]{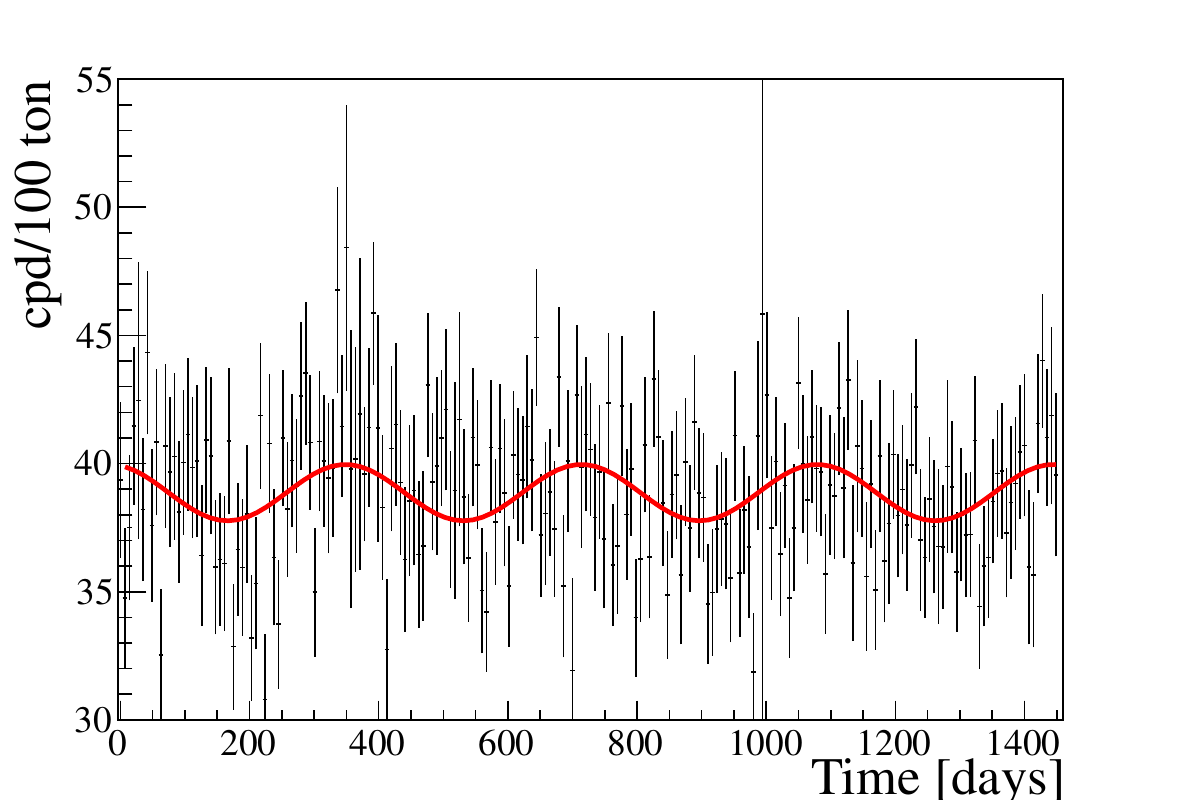}}
\caption{Rate of $\beta$-like events passing selection cuts with 7-day bins. The red line is the the result from Lomb-Scargle analysis (Eq.~\ref{equ:sinwave}.)}
\label{fig:R050}
\end{center}
\end{figure}

\begin{figure}[!h]
\begin{center}
\centerline{\includegraphics[width=1.0\linewidth, angle=0]{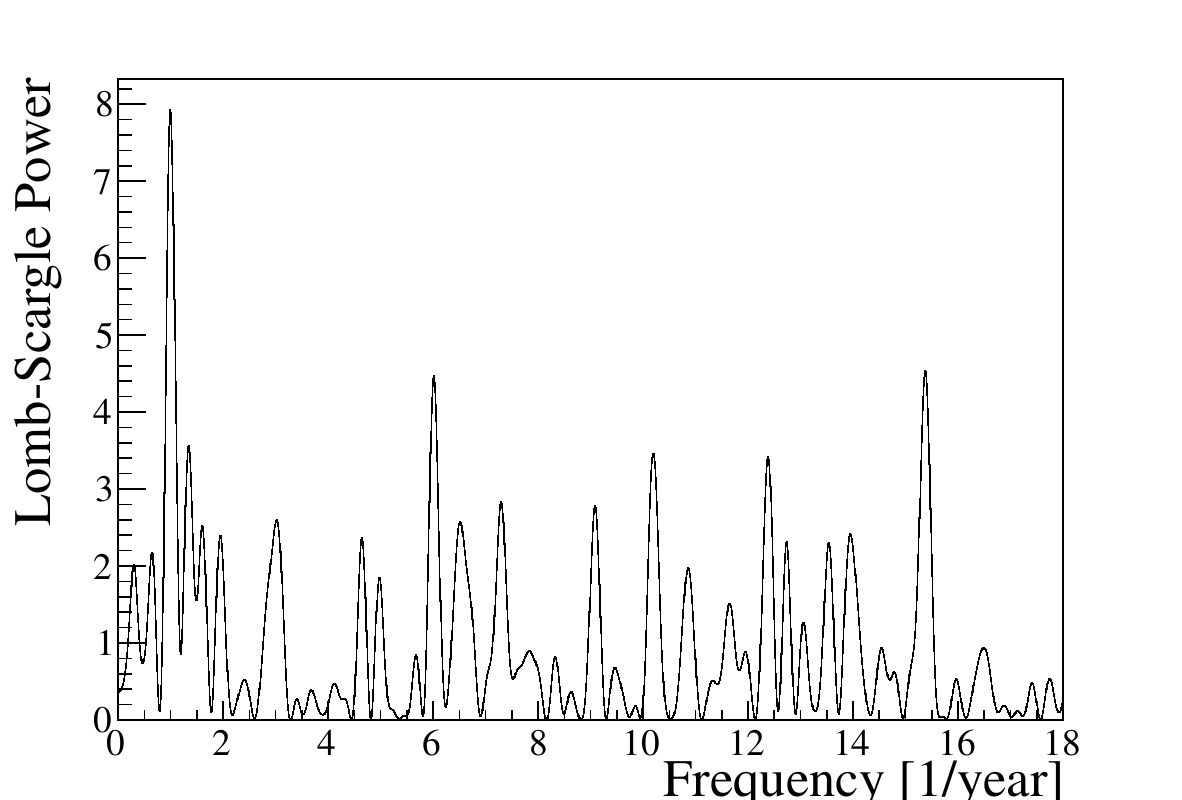}}
\caption{\small {Lomb-Scargle periodogram for data shown in Fig.~\ref{fig:R050}. 
}}
\label{fig:lombtot}
\end{center}
\end{figure}

For this analysis the data are grouped, after selection cuts, into 7-day bins as shown in Fig.~\ref{fig:R050}.
The Spectral Power Density $P(f)$ is calculated using the corresponding normalized event rate $R(t_{k})$
 and it is shown in Fig.\ref{fig:lombtot}. 

\begin{figure}[!h]
\begin{center}
\centerline{\includegraphics[width=1.\linewidth, angle=0]{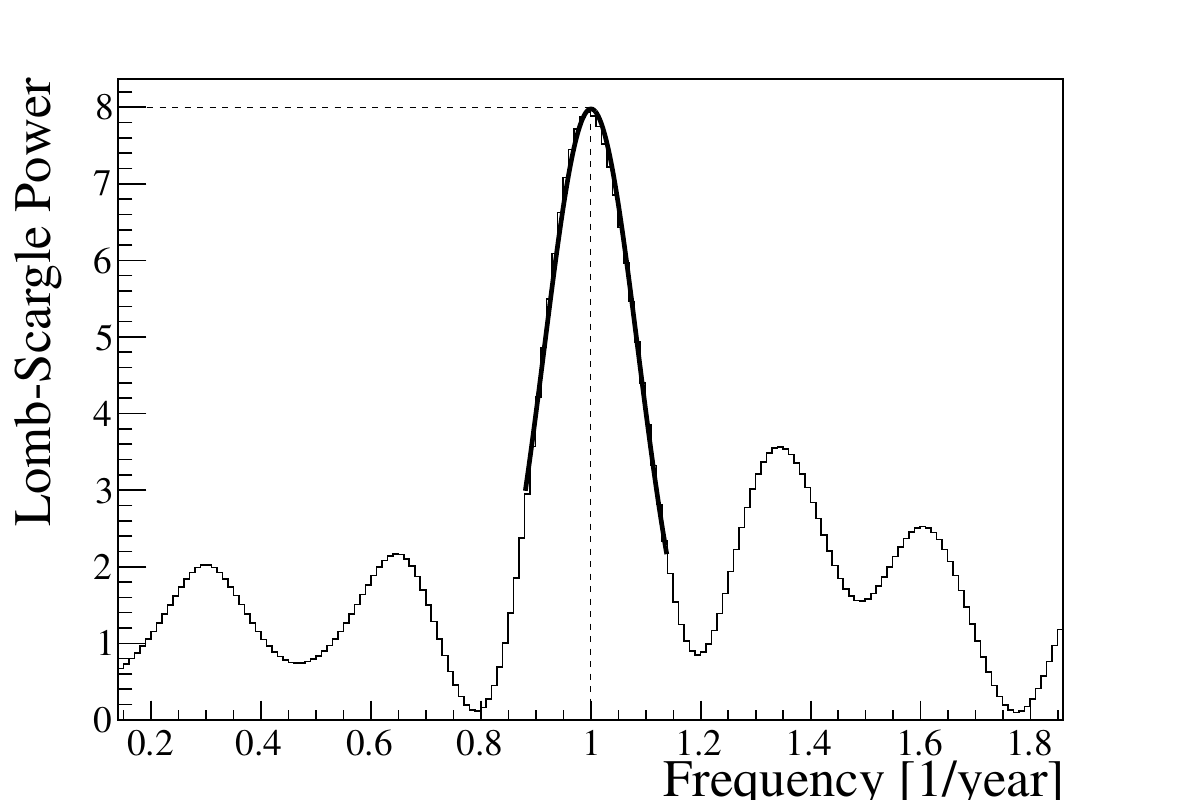}}
\caption{\small{Zoom-in of Fig.~\ref{fig:lombtot}. The peak $P(f)$ (1-year) is identified to be 7.9, as indicated by the vertical line.
}}
\label{fig:lomb105}
\end{center}
\end{figure}

The maximum of the periodogram is at $f=1~\mathrm{year}^{-1}$ and corresponds to  a $P(f)$  value of 7.9. A zoom-in is shown in Fig.~\ref{fig:lomb105}.

\begin{figure}[!h]
\begin{center}
\centering{\includegraphics[width = 0.5 \textwidth, angle=0]{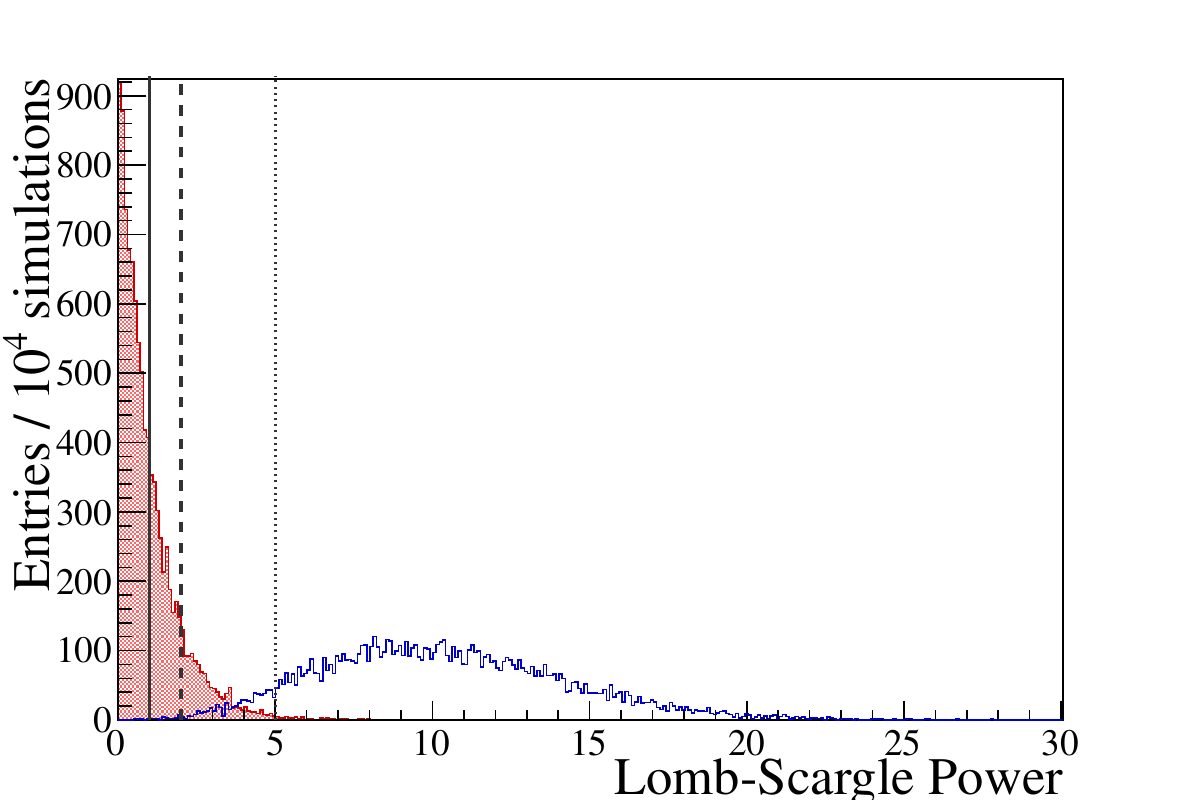}}
\caption{Detection sensitivity. Distributions of the Lomb-Scargle Power  at frequency corresponding to a 1 year period for $10^{4}$ simulations of a 6.7\% solar neutrino annual flux modulation with constant background (blue line) and the same number of white-noise simulations (background without any signal) (red area). Indicated with vertical lines are the sensitivity thresholds of 1$\sigma$ (solid), 2$\sigma$ (dashed), and 3$\sigma$ (dotted) C.L. above the white noise.
}
\label{fig:LSMC}
\end{center}
\end{figure}

Following ~\cite{bib:horne}, we have evaluated the significance of the largest peak found in the periodogram of our experimental data set with a toy Monte Carlo simulation assuming a realistic signal-to-background ratio and a  time interval of 4 years. 
Figure~\ref{fig:LSMC} displays the $P(f)$, at $f = 1~\mathrm{year}^{-1}$, distribution (red filled area) obtained applying the Lomb-Scargle analysis to $10^{4}$ simulations of a constant rate signal corresponding to the null
hypothesis (absence of modulation). This distribution is exponential as expected
for the power at a given frequency of the standard Lomb-Scargle periodogram of a pure white noise time series, $Prob(P(f)>z)= e^{-z}$~\cite{bib:Lomb, bib:Scargle, bib:gio}. 
In the plot, the vertical lines mark the $1\sigma$ (solid), $2\sigma$ (dashed) and $3\sigma$ (dotted) sensitivity to the null hypothesis. The blue distribution is obtained from $10^{4}$ simulations of an expected yearly modulated signal plus constant backgrounds and its most probable value is $P(f)=9.9$ with $rms$ of 4.

The Spectral Power Density $P(f)$ of 7.9 for $f = 1~\mathrm{year}^{-1}$, obtained from the data, is within the range expected from Monte Carlo and corresponds to $>3.5\sigma$ significance with respect to the null hypothesis. 

In addition we have estimated via Monte Carlo the significance of the two 4.5 high peaks
in the L-S periodogram.  Missing any a-priori information about the presence of periodicities
other than the annual one, the significance of these two peaks must be evaluated as
global significance, which takes into account the so called Look Elsewhere Effect,
i.e. the blind search over a frequency range \cite{bib:gio}. Basically, one performs a Monte Carlo evaluation of the distribution of the highest peak induced by a pure noise time series over the
searched frequency interval. The significance (or p-value) is computed comparing the obtained distribution with the Power value of the highest peak detected in the Lomb-Scargle periodogram of the data. In this way  we determined for  the two 4.5 high peaks the p-value of 85\%. Hence these two peaks are fully compatible with being pure noise induced fluctuations in the spectrum.

Finally, a sinusoidal function is constructed via Eq.~(\ref{equ:sinwave}) for $f_0=1~\mathrm{year}^{-1}$ and overlaid to the time-binned data in Fig.~\ref{fig:R050} (red curve).
The peak-to peak amplitude is $\sim 5.7\%$, slightly less than that expected from the eccentricity of the Earth's orbit, because the Lomb-Scargle method cannot disentangle the background from neutrino signal.
The same analysis using data selected with slightly different cuts and without applying the rate correction for MLP inefficiency (see Sec.~\ref{sec:mlpeff}), returns consistent results. The resulting total uncertainty for the period is $4\%$, and for the amplitude $7\%$. No phase information is available with this technique.

\subsection{Empirical Mode Decomposition}
\label{sec:EMD}

The third method, the {\it ``Empirical Mode Decomposition''} (EMD) \cite{bib:Huang98,bib:WuHuangEEMD2009}, has been
designed to work with non periodical signal, in order to extract the main parameters
from a time series as instantaneous frequency, phase and amplitude. 
The algorithm does not make any assumption about the functional form of the signal, in contrast to the 
Fourier analysis, and can therefore extract any time variation embedded in the data set.   

The EMD is a methodology developed to perform time-spectral analysis based on a empirical and iterative algorithm called
{\it sifting}, able to decompose an initial signal in a set of complete, but not orthogonal, oscillation mode 
functions called {\it "Intrinsic Mode Function"} or IMF \cite{bib:WuHuang2010}. 

Here we adopt a new technique for the noise assisted method called {\it ``Complete Ensemble Empirical Mode Decomposition with Adaptive Noise''} 
(CEEMDAN) \cite{bib:Torres2011} showing a greater efficiency and stability on the final results than the EEMD method \cite{bib:long}.
The algorithm is more capable to separate the signals of interest from background because it removes the residual 
noise present in the final IMFs together with the spurious oscillation modes \cite{bib:Colominas2014}.

\subsubsection{Standard Algorithm}

\begin{figure}[h!]
\centerline{\includegraphics[width=0.5 \textwidth, angle=0]{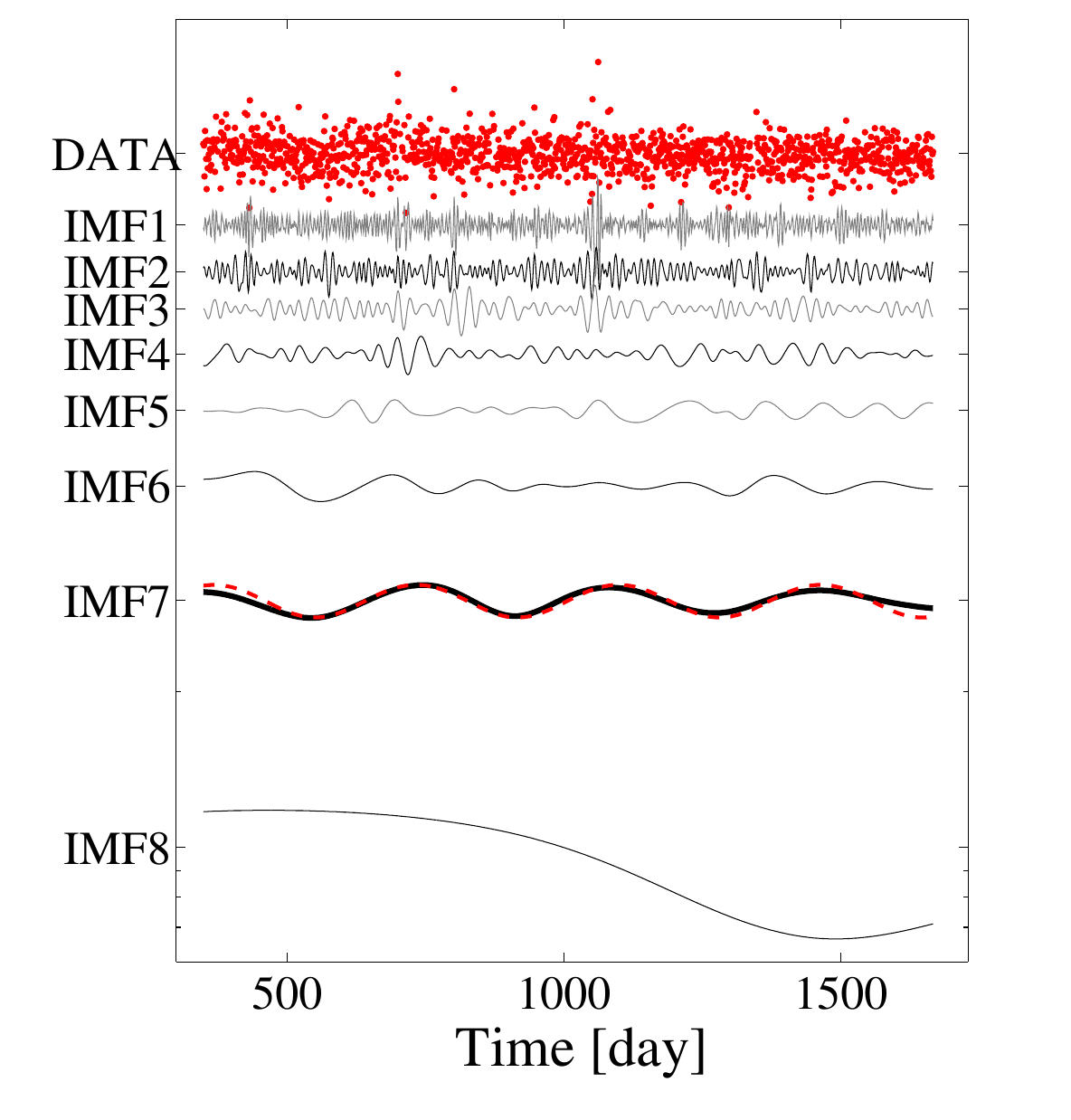}}
\caption{\small{Results of the {\it sifting} algorithm. The red points are the data 1-day binned. The $y$-axis is in log scale to show the shapes of the last IMFs that are too small with respect to the first ones.
The IMF-7 (solid black line) is compatible with the expected signal from Seasonal Modulation (dashed red line) and the last IMF is the trend.}}
\label{fig:SiftingRes}
\end{figure}

The {\it sifting} algorithm (Sec. \ref{sec:EMD}) requires a large number of points for a best performance. 
To maximize this number we chose bins of 1 day. As a consequence, statistical fluctuations dominate the dataset time-series
(red points in Fig.~\ref{fig:SiftingRes}). However, the intrinsic {\it dyadic} filter~\cite{bib:WuHuang2004}, 
removes all high frequency components created by the Poisson statistical noise.

The intrinsic mode functions, IMFs, are extracted from the original function through an iterative
procedure: the {\it sifting} algorithm. The basic idea is to interpolate at each step the local maxima and
minima of the initial signal, calculate the mean value of these interpolating functions, and
subtract it from the initial signal. The same procedure is then repeated on the residual subtracted signal
until suitable stopping criteria are satisfied. 
These are numerical conditions, which slightly differ in literature according to the approach followed (see {\it e.g.}~\cite{bib:Huang98,bib:Rilling2003}).
They aim at making sure that the IMFs obey two features inherited from harmonic functions: first, the number of extrema (local maxima
and minima) has to match the number of zero crossing points or differ from it at most by one;
second, the mean value of each IMF must be zero. 

The $i$-th IMF obtained by the $k$-iteration is given by:

\begin{equation} 
	IMF_{i}(t) = x_{i}(t) - \sum_{j=1}^{k}m_{ij} 
\end{equation}

where $x_{i}(t)$ is the residual signal when all ``i-1'' IMF's have been subtracted from the
original signal $R(t)$, $x_{0}(t) = R(t)$, and the $m_{ij}$ are the average function of the max and min
envelopes at each $j$-th iteration. Following the results from a detailed simulation, 
we fixed the number of sifting iterations to 20. This number guarantees a good
symmetry of the IMF with respect to its mean value, preserving the dyadic-filter property of the
method ({\it i.e.}, each IMF has an average frequency that is half of the previous one
\cite{bib:WuHuang2010}). Thus we obtain all $i$ IMFs down to the last one called
{\it ``trend''}, that is a monotonic IMF. 

The EMD approach features two potential issues: on one hand, the method is strongly dependent on small changes of
the initial conditions; on the other, mode mixtures could occur for a physical component present in the
data set especially when the ratio between signal and noise\footnote{In this case noise
means the statistical fluctuations of the rate with respect to the amplitude of the seasonal
modulation signal.} is low (about $S/N=0.2$, in our case). In order to account for these problems, a noise-assisted 
technique has been adopted. A random white noise signal (dithering) was added several times to
the data set under study and the average of all the IMFs taken.

As for the Borexino Phase-I analysis \cite{bib:long}, we repeat the single extraction of the IMF
1000 times, adding to the data a white noise component with an average value $\mu_{wn} = 0.0$ and 
$\sigma_{wn} = \sqrt{N_{bin}}$, where $N_{bin}$ is the rate of the single bin (Poisson's error).
The main difference with respect to the Borexino Phase-I analysis is the use of the noise-assisted approach, 
called CEEMDAN. 

The final decomposition of our data set is shown in Fig.~\ref{fig:SiftingRes}, where the lower frequency components identified by the algorithm become visible in the higher IMFs.
The ones shown are the resulting IMFs averaged over the 1000 extractions with different regenerations of 
white noise.

In particular, Fig.~\ref{fig:IMFp2bc}c shows the grey band corresponding to 1000 noise regenerated
IMF-7 containing the seasonal modulation. The resulting average
function is shown as black solid line, while the red-dashed curve corresponds to the expected seasonal modulation.

\subsubsection{Modulation Parameters Estimation}
Here we can only provide a short account of the procedures to calculate the modulation parameters. A more
detailed and formal description of the numerical calculations and theoretical explanations are
reported in \cite{bib:Huang98,bib:onIF}.

The frequency and the amplitude values of a periodic function (as the seasonal modulation) are
constant in time. We therefore expect that in the IMF7 (Fig.~\ref{fig:SiftingRes}) where a 
modulation of 1-year period is visible, these parameters will be constant in time, the average curve peaking on the 
expected values.
Naturally, due to the numerical procedure with which the {\it ``signal''} has been obtained, some
small fluctuations of the frequency and of the amplitude are expected.

\begin{figure}[!t] 
\begin{center}

\centerline{
\includegraphics[width=0.5\textwidth, angle=0.0]{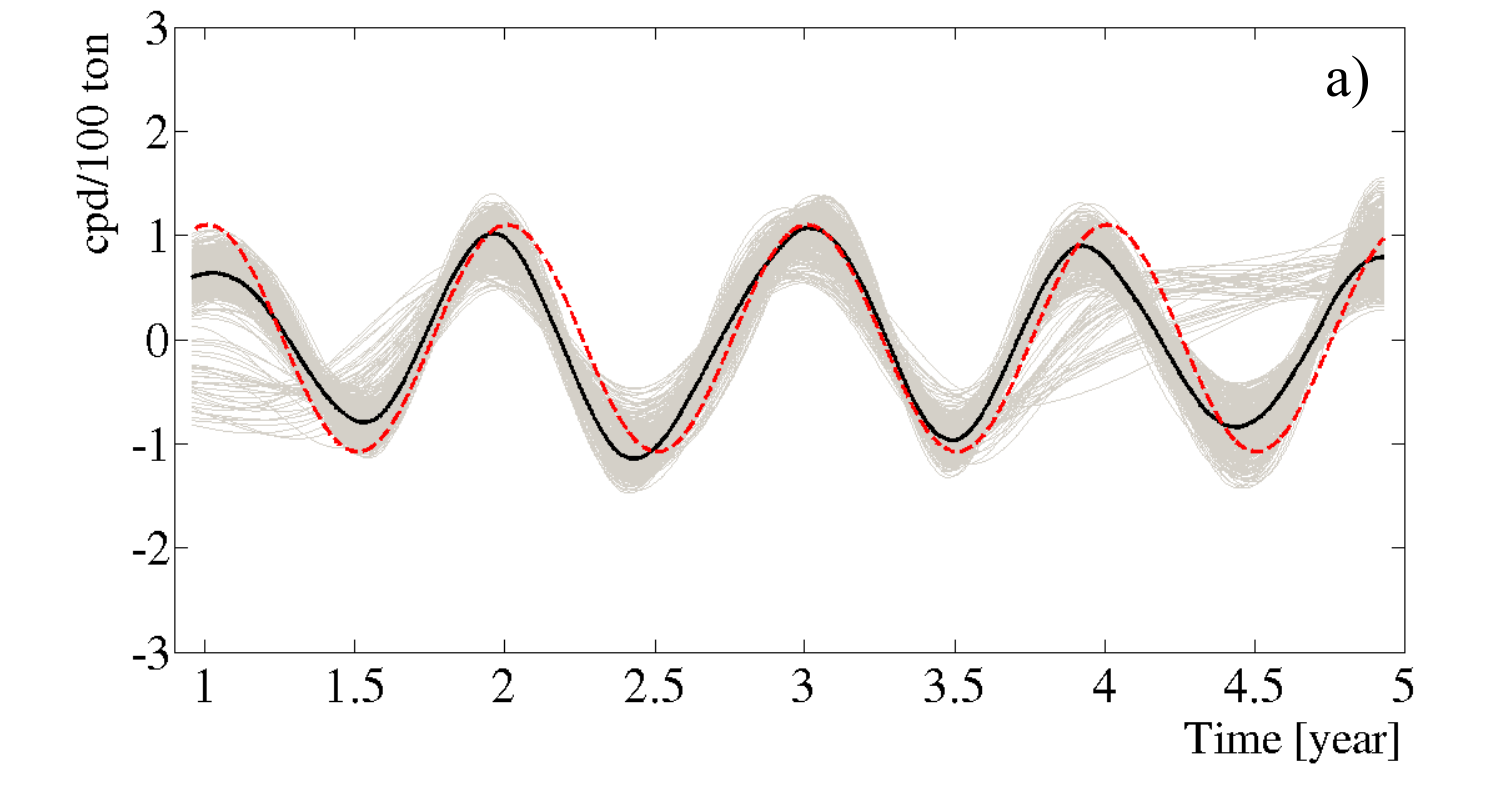}}
\centerline{\includegraphics[width=0.5\textwidth, angle=0.0]{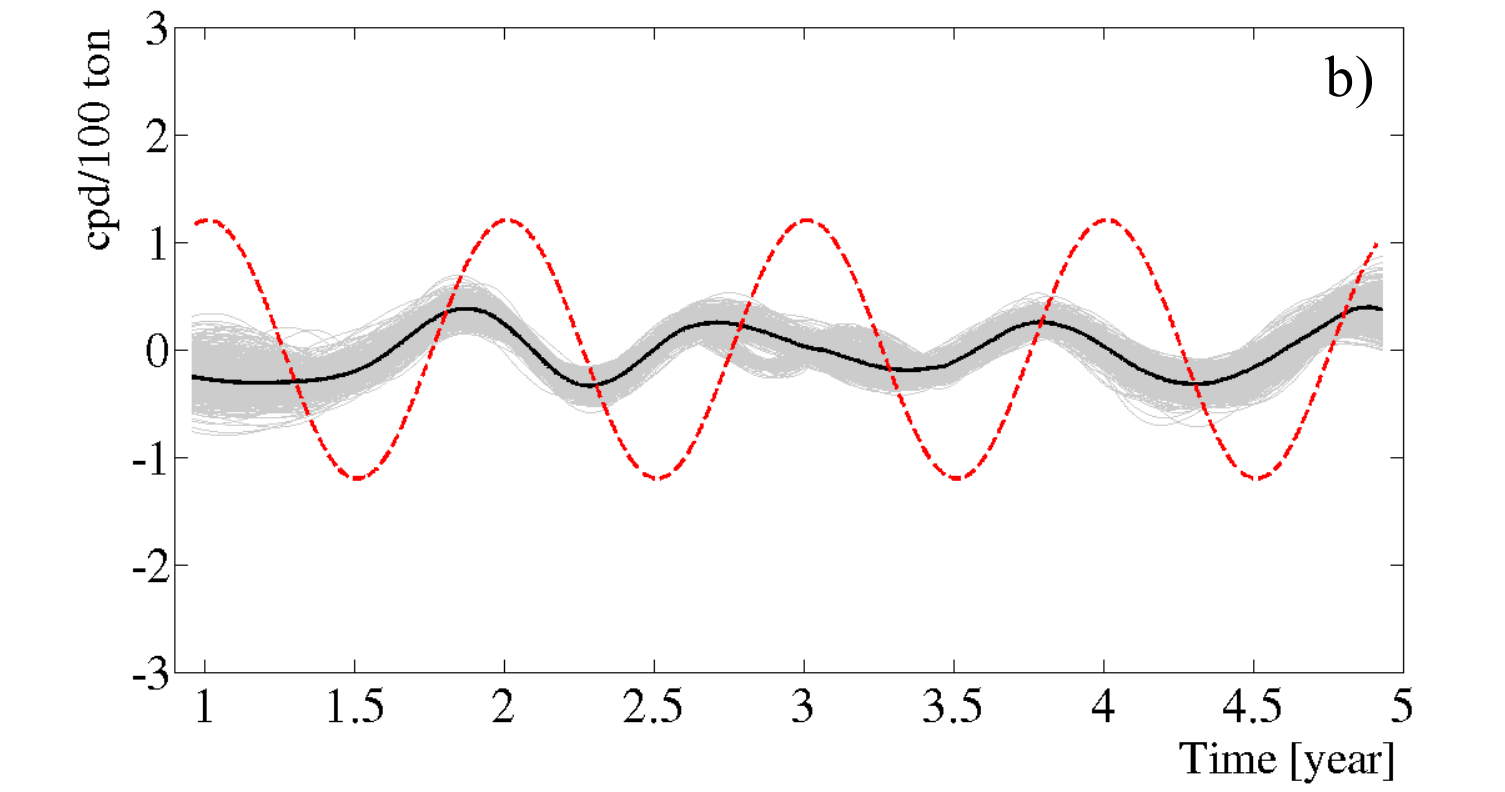}}
\centerline{\includegraphics[width=0.5\textwidth, angle=0.0]{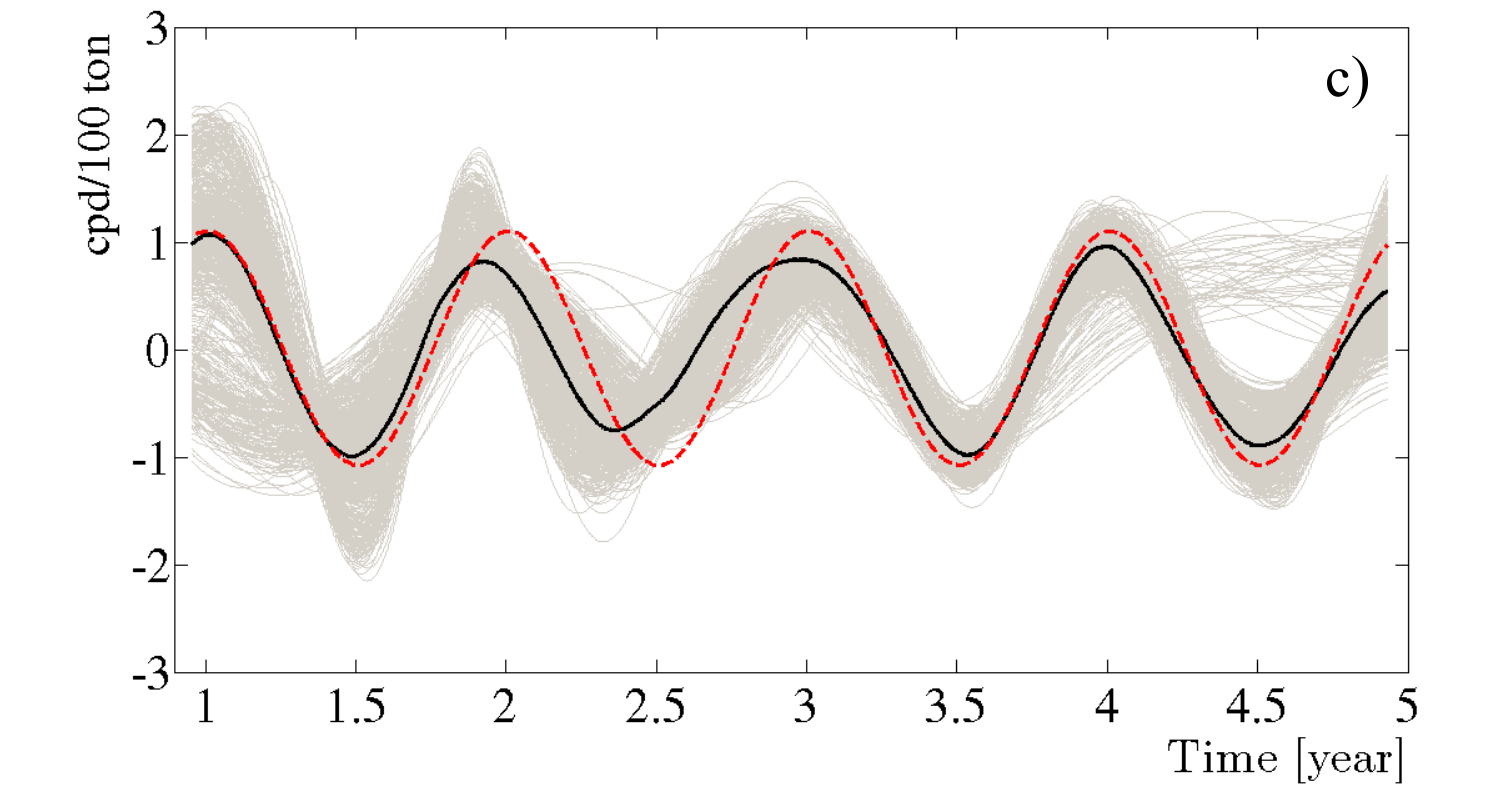}}
\caption{\small{In figure (a), (b) and (c) are shown in grey the 1000 functions collected for the IMF7 
and in black the relative average for the three different datasets: (a) simulated
seasonal modulation, (b) simulated constant background and (c) the real data set (see Fig.~\ref{fig:SiftingRes}). The red-dashed line is the expected seasonal modulation.}}
\label{fig:IMFp2bc}
\end{center}
\end{figure}

The IMF functions extracted by the {\it sifting} algorithm are not based on an analytical function.
Therefore, in order to extract information on frequency, phase and amplitude, it is necessary to build a
complex function $z(t)$ by means of a Hilbert transform of the initial signal  \cite{bib:Huang98}:

\begin{equation} 
z(t)=a(t)+\mathrm{i}b(t)=A(t) e^{- \mathrm{i} \theta(t)},
\label{eq:anal_func}
\end{equation}
in which the real part $a(t)$ is the IMF and the imaginary part $b(t)$ is the 
Hilbert transform of the real function:

\begin{equation}
b(t)= \frac{1}{\pi}P\int_{t'} \frac{a(t')}{(t-t')} dt'
\end{equation}
where $P$ is the Cauchy principal value. In Eq.(\ref{eq:anal_func}), $A(t)$ is defined as 

\begin{equation}
A(t)=\sqrt{a^2(t)+b^2(t)}.  
\label{eq:AmplCalc}
\end{equation}
A(t) is also called the amplitude modulation function (AM), while

\begin{equation}
\theta(t)=\arctan \left( \frac{b(t)}{a(t)}\right)
\label{eq:PhaseCalc}
\end{equation}
defines the phase of the carrier function or frequency modulation (FM) function.
This method provides a function of the phase of the time that we can use to define the
{\it instantaneous frequency} (IF) as simple time derivative of the phase $\theta(t)$.
Unfortunately a direct calculation of the IF, starting from the signal, 
gives unphysical results with negative values for the frequencies. In order to solve this problem, 
an additional numerical procedure is required: the {\it``Normalized Hilbert Transform''} (NHT) 
\cite{bib:onIF}. 
Performing the NHT we obtain a normalized carrier function over all the time series.
Building the $z(t)$ function, we are able to calculate a reliable instantaneous frequency 
function with a real physical meaning as follows:

\begin{equation}
f(t)=\dfrac{{\rm d}\theta(t)}{{\rm d}t}.
\label{eq:IstFrqCalc}
\end{equation}

We calculate the IF $f(t)$ and the amplitude $A(t)$ for all the IMFs extracted from each noise
regeneration and take the distribution of their average in time. A
Gaussian fit is applied to the resulting distribution to obtain $f$(t), A(t) and their respective errors.
  
In Fig.~\ref{fig:IMFp2bc} we compare IMFs obtained from the real dataset (Fig.~\ref{fig:IMFp2bc}c) with simulated data sets from a toy Monte Carlo with/without the sinusoidal signal expected for the seasonal modulation (Fig.~\ref{fig:IMFp2bc}a and \ref{fig:IMFp2bc}b respectively).

For both real and MC data set, the resulting IMF average shows a very good agreement with the expected seasonal modulation function, while in the case of the null hypothesis (Fig.~\ref{fig:IMFp2bc}b) the 
amplitudes of the resulting IMFs are substantially smaller while frequencies and phases are varying randomly. 

\begin{figure}[!t]
\begin{center}

\centerline{ \includegraphics[width=0.5\textwidth,angle=0.0]{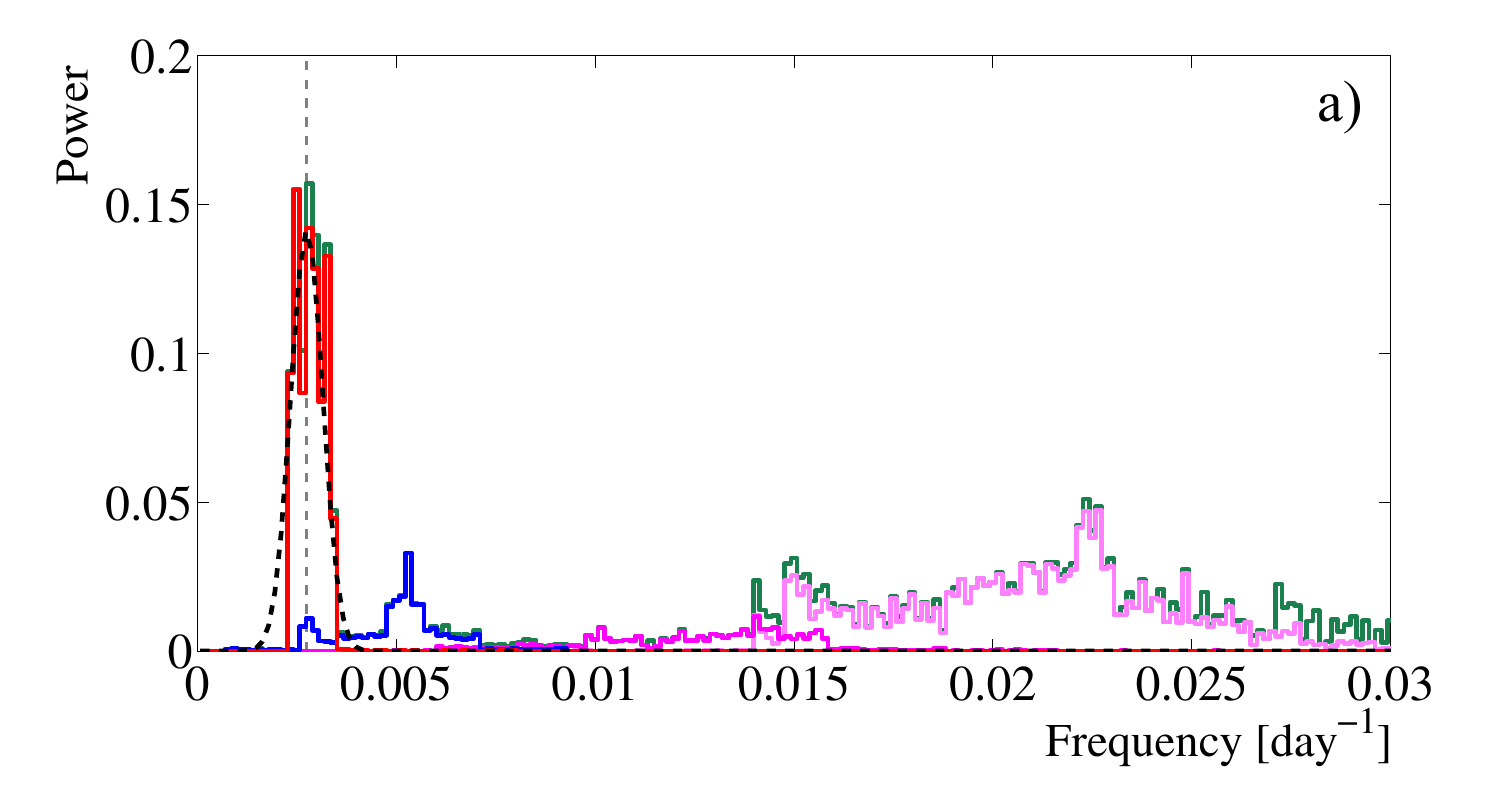}}
\centerline{ \includegraphics[width=0.5\textwidth,angle=0.0]{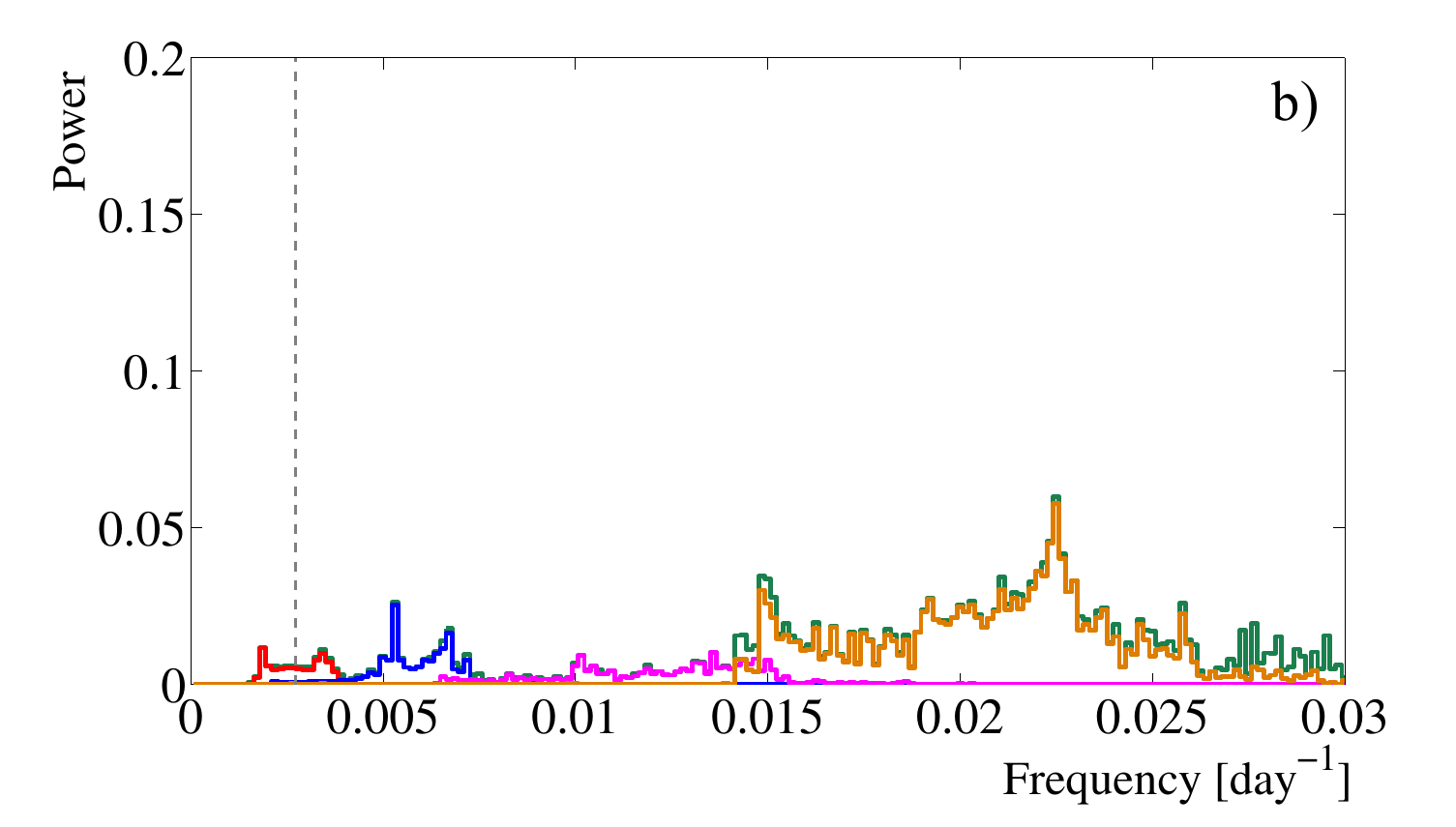}}
\centerline{\includegraphics[width=0.5\textwidth,angle=0.0]{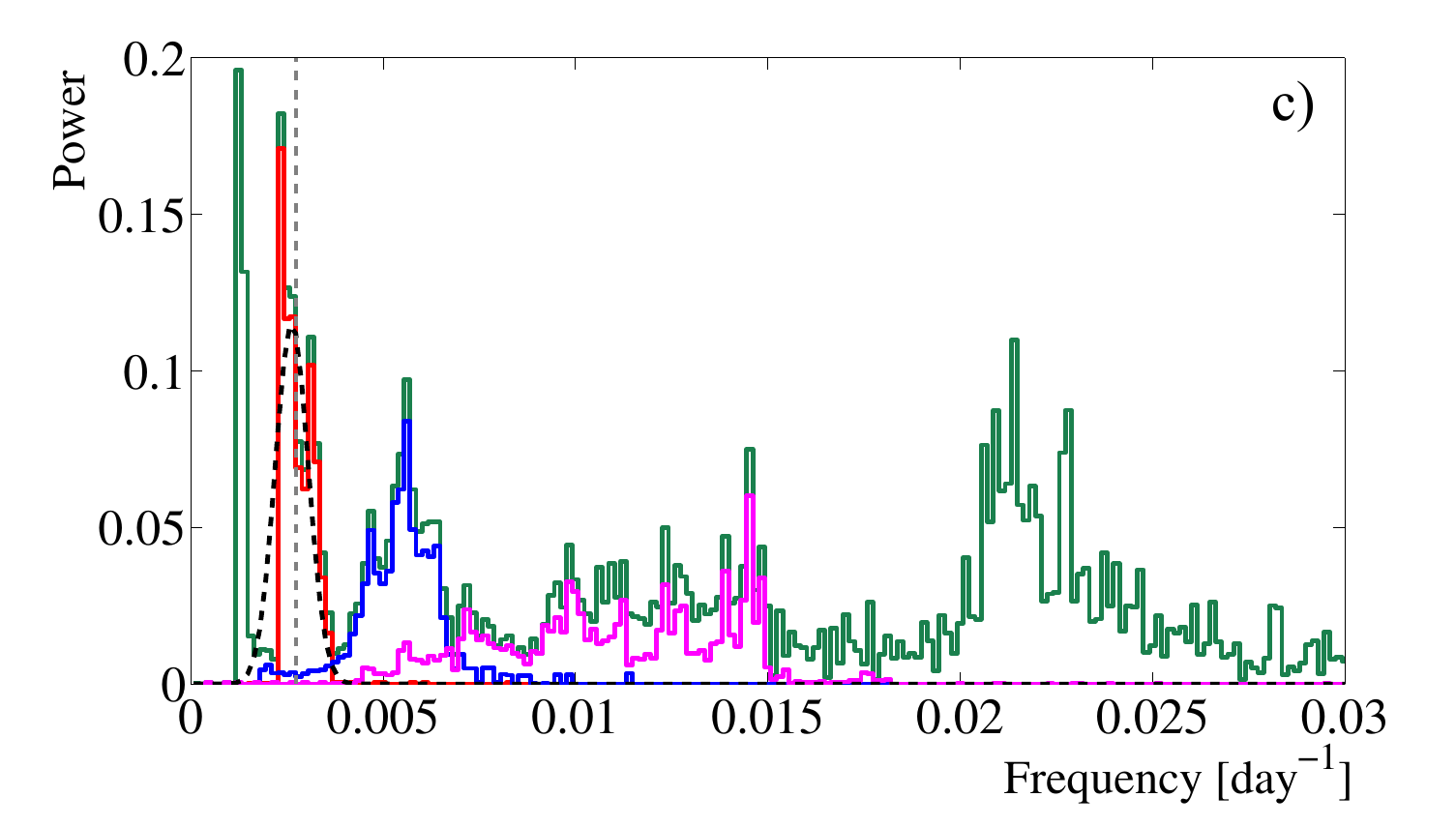}}
\caption{\small{Power spectrum ($\langle A^2(t)\rangle$) (Eq. \ref{eq:AmplCalc}). The dark-green solid line is the power spectrum of the full data set,
while the other colored spectra represent the components from the last 4 IMFs.
The red solid line is the power spectrum of IMF-7, where the seasonal modulation is present. (a) simulated seasonal modulation, (b) simulated constant background and (c) the real data set.
}}
\label{fig:IMF_Power}
\end{center}
\end{figure}

A power spectrum is defined based on the average in time of the square amplitudes 
($\langle A^2(t)\rangle$) (\ref{eq:AmplCalc})  for each frequency $\omega (t)$.
Fig.~\ref{fig:IMF_Power} shows the relative power spectra for the simulations with and without modulation
(Fig.~\ref{fig:IMF_Power}a and ~\ref{fig:IMF_Power}b) and the real data set, respectively (Fig.~\ref{fig:IMF_Power}c).

The colored histograms are the Power Spectra from the last 4 IMFs, while the dark green 
are the full spectra of the whole set of IMFs (full dataset spectrum).

As expected in the presence of the seasonal modulation signal (Fig.~\ref{fig:IMF_Power}a and \ref{fig:IMF_Power}c), we observe a
narrow peak centered on the expected frequency ($f=1/T  = 2.73 \times 10^{-3} \;\mathrm{day}^{-1}$), while in the case of the null hypothesis this spectral component remains 
almost flat, featuring an amplitude comparable with other background IMFs that are present at higher 
frequencies. The power is an order of magnitude lower than the signal case (Fig.~\ref{fig:IMF_Power}b).

Applying equation~\ref{eq:IstFrqCalc}, we compute the average parameters shown in Tab.\ref{tab:finalRes} for the simulated and real data. The results are in agreement with the expected seasonal modulation. 

\begin{table}[!h]
\begin{center}
\begin{tabular}{l|c|c}

\hline\hline
              & Simulated Data                              & Data     \\
\hline        

$T$ [year]        & $0.95 \pm 0.02$                 & $0.96 \pm 0.05$   \\
$\varepsilon$ & $0.0155 \pm 0.0025 $           &  $0.0168 \pm 0.0031 $    \\
$\phi$ [day]   & $-12 \pm 11$                          & $14 \pm 22$ \\ 
\hline
\end{tabular}

\caption{\small{Period, eccentricity and phase of the solar neutrino seasonal modulation flux. The results
from data are in agreement with the Monte Carlo results.}}
\label{tab:finalRes}
\end{center}
\end{table}

Based on the comparison of the power spectrum and the parameters resulting from the zero-modulation MC data sets we conclude the presence of a seasonal modulation.

\begin{figure}[!t]
\begin{center}
\centerline{\includegraphics[width = 0.47 \textwidth, angle=0]{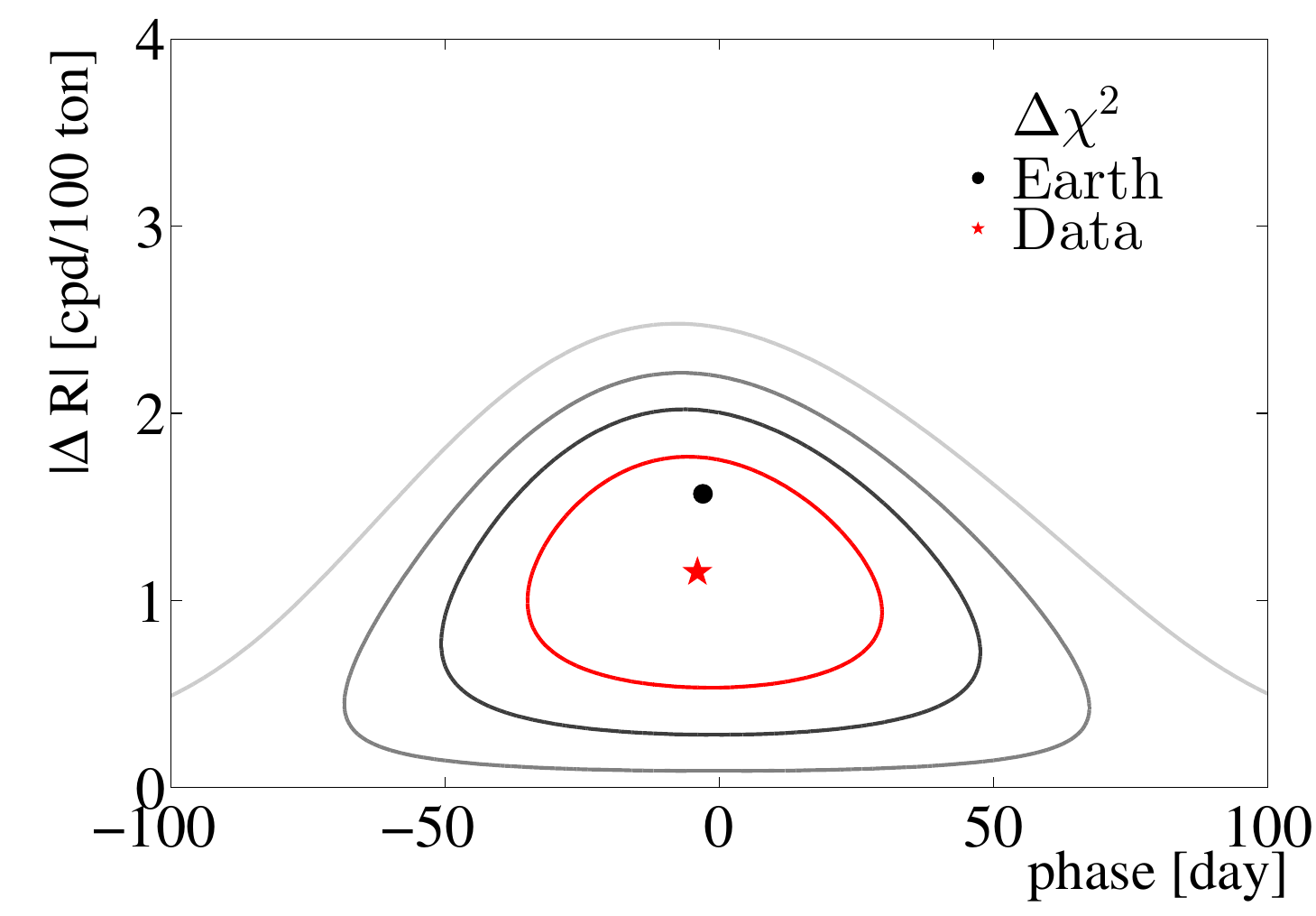}}
\caption{\small{Amplitude and phase obtained with the EMD method. The red star indicates the best-fit results,
while the black point the expected values. Confidence contours of 1, 2, and 3$\sigma$ are indicated with solid lines.}
\label{fig:EMDpar}}
\end{center}
\end{figure}

We have calculated a $\chi^2$-map varying both the phase and modulation amplitude of the sinusoidal function with respect to the average IMF obtained over the complete 1000 noise regenerations. The $\chi^2$-contours are displayed in Fig.~\ref{fig:EMDpar}, where we assumed the standard deviation of the IMFs from the average curve to equal 1$\sigma$-uncertainties divided by the number of time bins minus one.

\section{Summary}
\label{sec:end}

Four years of Borexino Phase-II data have been analyzed searching for the expected annual modulation of the $^7$Be solar neutrino interaction rate induced by the eccentricity of the Earth's orbit around the Sun.

Both the detector and the data have shown remarkable stability throughout the entire Phase-II period, allowing for the clear emergence of the annual periodicity of the signal.

Three analysis methods were employed: an analytical fit to event rate, a Lomb-Scargle periodogram and an Empirical Mode Decomposition analysis. Results obtained with all three methods are consistent with the presence of an annual modulation of the detected $^7$Be solar neutrino interaction rate.  Amplitude and phase of the modulation are consistent with that expected from the eccentric revolution of the Earth around the Sun, proving the solar origin of the low energy neutrinos detected in Borexino.
The absence of an annual modulation is rejected with a 99.99\% C.L.. The direct fit to the event rate yields an eccentricity of $\epsilon = (1.74 \pm 0.45)\%$, while the Lomb-Scargle method identifies a clear spectral maximum at the period T=1 year. The EMD method provides a powerful and independent confirmation of these results.

\section*{Acknowledgements}

The Borexino program is made possible by funding from
INFN (Italy), NSF (USA), BMBF, DFG, HGF and MPG
(Germany), RFBR (Grants 16-02-01026 A, 15-02-02117 A, 16-29-13014 ofi-m,
17-02-00305 A) (Russia), and NCN Poland (Grant No. UMO-2013/10/E/ST2/00180). We acknowledge the generous hospitality and support of the Laboratory Nazionali del Gran Sasso (Italy).


\begin{thebibliography}{00}
\section*{References}

\bibitem{bib:detector} {
G. Alimonti et al. (Borexino Collaboration), The Borexino detector at the Laboratori Nazionali del Gran Sasso, Nucl. Instr. and Methods A 600 (2009) 568.}
\bibitem{bib:be7} {G. Bellini et al., (Borexino Collaboration), Precision Measurement of the 7Be Solar Neutrino Interaction Rate in Borexino, Phys. Rev. Lett. 107 (2011) 141302.}
\bibitem{bib:pep} {Bellini, G. et al. (Borexino Collaboration),
First Evidence of pep Solar Neutrinos by Direct Detection in Borexino, Phys. Rev. Lett. 108, (2012) 051302.}
\bibitem{bib:bxpp} {
G. Bellini et al. (Borexino Collaboration), Neutrinos from the primary  proton-proton  fusion  process  in  the  Sun,  Nature  512 (2014) 383.}
\bibitem{bib:b8} {G. Bellini, et al. (Borexino Collaboration), Measurement of the solar 8B neutrino rate with a liquid scintillator target and 3 MeV energy threshold in the Borexino detector, Phys. Rev. D 82, (2010) 033006.}
\bibitem{bib:carlos} {J. Bergstršm, M.C. Gonzalez-Garcia, M. Maltoni et al, Updated determination of the solar neutrino fluxes from solar neutrino data, J. High Energ. Phys. (2016) 132.}
\bibitem{bib:geo}  {M. Agostini et al. (Borexino Collaboration),  Spectroscopy of geoneutrinos from 2056 days of Borexino data, Phys. Rev. D 92, 031101(2015) 031101.}
\bibitem{bib:anti}  {G. Bellini, G. et al. (Borexino Collaboration), Study of solar and other unknown anti-neutrino fluxes with Borexino at LNGS, Phys. Lett. B 696, (2011) 191.}
\bibitem{bib:muon} {G.  Bellini  et  al.  (Borexino  Collaboration),  Muon  and  Cosmogenic  Neutron  Detection  in  Borexino,  J.  Instrum.  6  (2011) P05005.}
\bibitem{bib:serenelli16} { A. Serenelli, Alive and well: a short review about standard solar models, 
arXiv: hep-ph/1601.07179 (2016)}
\bibitem{bib:be2007} Bellini, G. et al. (Borexino Collaboration), First real time detection of 7Be solar neutrinos by Borexino, Phys. Lett. B 658  (2008) 101.
\bibitem{bib:long} {
G.  Bellini  et  al.  (Borexino  Collaboration),  Final  results  of Borexino  Phase-I  on  low-energy  solar  neutrino  spectroscopy, Phys. Rev. D 89 (2014) 112007.}

\bibitem{bib:SNO} {B. Aharmim (SNO Collaboration), Search for periodicities in the B8 solar neutrino flux measured by the Sudbury Neutrino Observatory, Phys. Rev. D 72 (2005) 052010.}
\bibitem{bib:SK} {J. Hosaka et al. (Super-KKamiokande Collaboration), Solar neutrino measurements in Super-Kamiokande-I, Phys. Rev. D 73, (2006) 112001.} 

\bibitem{bib:homestake} {R. Davis Jr.,  A review of measurements of the solar neutrino flux and their variation, Nucl. Phys. B - Proceedings Supplements, 48, (1996) 284.}
\bibitem{bib:mils} {A.Milsztajin, A Search for Periodicity in the Super-Kamiokande Solar Neutrino Flux Data
arXiv:hep-ph/0301252 (2003).}
\bibitem {bib:stu} {D.O. Caldwell and P.A. Sturrock, Further Evidence for Neutrino Flux Variability from Super-Kamiokande Data, arXiv:hep-ph/0305303 (2003)}
\bibitem {bib:stu2} {D.O. Caldwell and P.A. Sturrock, Evidence for solar neutrino flux variability and its implications, Astropart. Phys., 23, (2005) 543.}
\bibitem{bib:SK1} {J.Yoo et al. (SK Collaboration), Search for periodic modulations of the solar neutrino flux in Super-Kamiokande-I, Phys.Rev. D68, (2003) 092002.}
\bibitem{bib:GNO} {L. Pandola, Search for time modulations in the Gallex/GNO solar neutrino data
Astropart. Phys. 22 (2004) 219.}

%analysis
%\bibitem{bib:paperMLP} Borexino Collaboration, to be submitted.

\bibitem{bib:Lomb} {N.R. Lomb, Least-squares frequency analysis of unequally spaced data, Astrophysics and Space Science, 39 (1976) 447.}
\bibitem{bib:Scargle} {J.D. Scargle, Studies in astronomical time series analysis. II - Statistical aspects of spectral analysis of unevenly spaced data, Astrophysical Journal 263 (1982) 835.}

\bibitem{bib:Huang98} {N.~E. Huang, Z. Shen, S.~R. Long et al., The empirical mode decomposition and the Hilbert spectrum for nonlinear and non-stationary time series analysis, Proc. R. Soc. of Lond. A,  454 (1998) 903.}

\bibitem{bib:MSW} {G. Bellini, G. et al. (Borexino Collaboration), Absence of a day-night asymmetry in the 7Be solar neutrino rate in Borexino, Phys. Lett. B 707, (2012) 22.}


%lomb
\bibitem{bib:horne}{J. H. Horne and S. L. Baliunas, A prescription for period analysis of unevenly sampled time series, Astrophys. J. 302 (1986) 757.}
\bibitem{bib:gio} {G.~Ranucci., Likelihood scan of the Super-Kamiokande I time series data,
Phys. Rev. D 73  (2006) 103003.}

% EMD
\bibitem{bib:WuHuang2010} {Z. Wu, N.E. Huang, On the filtering properties of the empirical mode decomposition, Advances in Adaptive Data Analysis, 2, No. 04, (2010) 397.}
\bibitem{bib:Torres2011}
{M.E. Torres, M.A. Colominas, G. Schlotthauer, P. Flandrin, A complete ensemble empirical mode decomposition with adaptive noise, Proc. of 36th IEEE  Intern. Conference on Acoustic, Speech and Signal Processing, Prague, Czech Republic, (2011) 4144. }
\bibitem{bib:Colominas2014} {M.A. Colominas, G. Schlotthauer G., M.E. Torres, Improved complete ensemble EMD: A suitable tool for biomedical signal processing, Biomedical Signal Processing and Control 14 (2014) 19.}
\bibitem{bib:WuHuangEEMD2009}
{Z. Wu and N. E. Huang, Ensemble empirical mode decomposition: a noise-assisted data analysis method, Advances in Adaptive Data Analysis, 1, No.1 (2009) 1.}


\bibitem{bib:WuHuang2004} 
{Z.Wu and N.E.Huang, A study of the characteristics of white noise using the empirical mode decomposition method, Proc. R. Soc. Lond. A 460 (2004) 1597.}


\bibitem{bib:Rilling2003} {G. Rilling, P. Flandrin, P. Goncalves, On empirical mode decomposition and its algorithms, Proc. of IEEE-EURASIP Workshop on Nonlinear Signal and Image Processing NSIP-03, (2003) 177.}

\bibitem{bib:onIF} {N.E.Huang et al., On instantaneous frequency,
Advanced in Adaptive Data Analysis,1, No.2, (2009) 177}



 \end{thebibliography}
\end{document}